\documentclass{aa}  

\usepackage{graphicx}
\usepackage{txfonts}
\usepackage{listings}
\usepackage{multirow}
\usepackage[dvipsnames]{xcolor}
\usepackage{array}
\usepackage[normalem]{ulem}
\usepackage{soul}
\usepackage{academicons}
\usepackage{xspace}

\newcommand{\code}[1]{\texttt{#1}\xspace}

\newcommand{\salt}{\code{SALT2}}
\newcommand{\saltmu}{\code{SALT2mu}}
\newcommand{\wfit}{\code{wfit}}
\newcommand{\bayes}{\code{StanIa}}

\newcommand{\snana}{\code{SNANA}}

\newcommand{\proj}[1]{\textsc{#1}\xspace}
\newcommand{\lsst}{\proj{LSST}}

\newcommand{\plasticc}{\proj{PLAsTiCC}}
\newcommand{\snpcc}{\proj{SNPhotCC}}
\newcommand{\resspect}{\proj{ReSSpect}}

\newcommand{\snia}{SN~Ia\xspace}
\newcommand{\sneia}{SNe~Ia\xspace}
\newcommand{\lc}{light curve\xspace}
\newcommand{\lcs}{light curves\xspace}

\newcommand{\fid}{\textit{fiducial}\xspace}
\newcommand{\ran}{\textit{random}\xspace}
\newcommand{\per}{\textit{perfect}\xspace}

\newcommand{\fomthree}{FOM$_{3}$\xspace}

\newcommand{\lcdm}{$\Lambda$CDM\xspace}

\newcommand{\revone}[2]{#2}%{{{\color{blue}\textit{#2}}}
\newcommand\Tstrut{\rule{0pt}{2.6ex}}         % = `top' strut
\newcommand\Bstrut{\rule[-0.9ex]{0pt}{0pt}}   % = `bottom' strut
\newcommand{\orcid}[1]{\href{https://orcid.org/#1}{\textcolor[HTML]{A6CE39}{\aiOrcid}}}

\usepackage[pagewise, switch]{lineno}
\begin{document} 

   \title{Are classification metrics good proxies \\
   for SN Ia cosmological constraining power?}
  \titlerunning{Evaluating classification metrics as \snia cosmology metrics}
%   \subtitle{I. Overviewing the $\kappa$-mechanism}

   \author{Alex I. Malz\inst{1}\thanks{\email{aimalz@nyu.edu}}
          \and
          Mi Dai\inst{2}
          \and
          Kara A. Ponder\inst{3}
          \and
          Emille E.O. Ishida\inst{4}
          \and
          Santiago Gonzalez-Gaitain\inst{5}
          \and
          Rupesh Durgesh\inst{6}
           \and
          Alberto Krone-Martins\inst{7,8}
          \and
          Rafael S. de Souza\inst{9}
          \and
          Noble Kennamer\inst{10}
          \and
          Sreevarsha Sreejith\inst{11}
          \and
          Llu\'is Galbany\inst{12,13}
          \and\\
          %\fnmsep\thanks{Just to show the usage
          %of the elements in the author field}
          The LSST Dark Energy Science Collaboration (DESC)
          \and
          the Cosmostatistics Initiative (COIN)
          }

   \institute{
   % Ruhr-University Bochum, Astronomical Institute, German Centre for Cosmological Lensing, 44801 Bochum, Germany
   %  \and
        McWilliams Center for Cosmology, Department of Physics, Carnegie Mellon University, Pittsburgh, PA, USA
       \and
       Department of Physics and Astronomy, Johns Hopkins University, Baltimore, MD 21218, USA
       \and
        SLAC National Accelerator Laboratory, 2575 Sand Hill Rd, Menlo Park, CA 94025, USA %\aim{ask for new affiliation?}
        \and
       Universit\'e Clermont Auvergne, CNRS/IN2P3, LPC, F-63000 Clermont-Ferrand, France
       \and
       CENTRA/COSTAR, Instituto Superior T\'ecnico, Universidade de Lisboa, Av. Rovisco Pais 1, 1049-001, Lisboa, Portugal
       \and
        Independent Researcher,  Ingolstadt, Germany %\emi{Look up company?}
        \and
        Donald Bren School of Information and Computer Sciences, University of California, Irvine, CA 92697, USA
       \and
       CENTRA/SIM, Faculdade de Ciências, Universidade de Lisboa, Ed. C8, Campo Grande, 1749-016, Lisboa, Portugal
       \and
       Centre for Astrophysics Research, University of Hertfordshire, College Lane, Hatfield, AL10~9AB, UK
        \and
         Department of Computer Science, University of California Irvine, Irvine, USA
        \and
        Physics Department, Brookhaven National Laboratory, Upton, NY 11973, USA
        \and
        Institute of Space Sciences (ICE, CSIC), Campus UAB, Carrer de Can Magrans, s/n, E-08193 Barcelona, Spain
        \and
        Institut d'Estudis Espacials de Catalunya (IEEC), E-08034 Barcelona, Spain
}

%   \date{Received September 15, 1996; accepted March 16, 1997}

% \abstract{}{}{}{}{} 
% 5 {} token are mandatory
 
  \abstract
   {When selecting a classifier to use for a supernova Ia (SN Ia) cosmological analysis, it is common to make decisions based on metrics of classification performance, i.e. contamination within the photometrically classified SN Ia sample, rather than a measure of cosmological constraining power.
   If the former is an appropriate proxy for the latter, this practice would save those designing an analysis pipeline from the computational expense of a full cosmology forecast.}
   {This study tests the assumption that classification metrics are an appropriate proxy for cosmology metrics.}
   {We emulate photometric SN Ia cosmology samples with controlled contamination rates of individual contaminant classes and evaluate each of them under a set of classification metrics. 
   We then derive cosmological parameter constraints from all samples under two common analysis approaches and quantify the impact of contamination by each contaminant class on the resulting cosmological parameter estimates.}
   {We observe that cosmology metrics are sensitive to both the contamination rate and the class of the contaminating population, whereas the classification metrics are insensitive to the latter.}
   {We therefore discourage exclusive reliance on classification-based metrics for cosmological analysis design decisions, e.g. classifier choice, and instead recommend optimizing using a metric of cosmological parameter constraining power.}

\maketitle
%-------------------------------------------------------------------

\section{Introduction}

Over twenty years after the discovery of the accelerating expansion of the universe \citep{Riess98,Perlmutter99}, Type Ia Supernovae (\sneia) remain a widely used probe of dark energy with the potential to distinguish between cosmological models and the values of their parameters, particularly the dark energy equation-of-state parameter, $w$.
Technological advances have allowed large photometric surveys such as the Sloan Digital Sky Survey\footnote{\url{https://www.sdss.org/}} (\proj{SDSS}), the \proj{ESSENCE} Supernova Survey \citep{vasey2007}, the SuperNova Legacy Survey \citep[\proj{SNLS}, ][]{astier2006},  \proj{Pan-STARRS}\footnote{\url{http://pswww.ifa.hawaii.edu/pswww/}} and the Dark Energy Survey\footnote{\url{https://www.darkenergysurvey.org/}} \citep[\proj{DES}, ][]{abbott2016}, and the Zwicky Transient Facility\footnote{\url{https://www.ztf.caltech.edu/}} \citep[\proj{ZTF}, ][]{bellm2019} 
to significantly increase the number of \sneia available for cosmological studies \citep{jones2018, Hlozek2012, popovic2020, vicenzi2022}. 
Soon the next-generation Rubin Observatory Legacy Survey of Space and Time\footnote{\url{https://www.lsst.org/}} (\lsst) and Nancy Grace Roman Space Telescope\footnote{\url{https://roman.gsfc.nasa.gov/}} will amass even larger samples of light curves, exceeding the available spectroscopic follow-up resources that could confirm their identities.
Consequently, the utility of these samples for \snia cosmology depends largely on the classification of their photometric \lcs \citep{Kessler10, Hlozek21}, primarily by machine learning techniques \citep[see e.g.~][ and references therein]{Ishida2019Nat}.

Since there is no perfect light curve classifier, we should expect
an unavoidable fraction of false positives (non-\sneia erroneously classified as \sneia), which can cause biases in subsequent cosmological analysis.
Imperfect classifications are in part induced by the coarseness of the broad-band photometry, the irregular and sparse timing of observations, and the non-representativity and incompleteness of training sets or model libraries. 
Much effort has been directed towards optimizing light curve classification, largely focusing on development of data-driven classifiers \citep[e.g. ][]{Muthukrishna2019,Pasquet2019, Moller2020, Villar2020} but also recently on improving training sets for machine learning methods \citep[e.g. ][]{Boone2019, Ishida19, Kennamer20, Carrick2021}. 
Valiant efforts toward using probabilistic classifications have been made \citep[e.g.~][]{KesslerScolnic2017}, yet the reliability of estimated classification probabilities remains difficult to characterize \citep{Malz19}, leading to a continued reliance on the definition of cosmological \lc samples with a goal of purity.

It is reasonable to expect that, depending on their \lc shape and color, each class of contaminant may induce a different biasing effect in the final cosmological results.
Thus, it is important not only to determine which classes of objects are the main sources of contamination, but also to understand how their contamination affects the cosmological results.
In what follows, we stress-test the hypothesis that metrics of classification quality are a good proxy for the impact of impurities on subsequent cosmological parameter inference.

This work was developed under the umbrella of the Recommendation System for Spectroscopic Follow-up (\resspect) project\footnote{\url{https://cosmostatistics-initiative.org/focus/resspect1/}}, a joint effort between the \proj{LSST} Dark Energy Science Collaboration\footnote{\url{https://lsstdesc.org/}} (DESC) and the Cosmostatistics Initiative\footnote{\url{https://cosmostatistics-initiative.org/}} (\proj{COIN}), whose goal is to guide the construction of optimal spectroscopic training sets for purely photometrically-typed \snia cosmology. 
The core project uses an active learning approach \citep[see e.g.,][]{Ishida19} that identifies, on each night, which objects should be selected for spectroscopic follow-up \citep{Kennamer20}.
Considering a fixed amount of telescope time per night there are different sets of objects which result in the same classification improvement if added to the training sample, however metrics of cosmological parameter constraints might be sensitive to effects that the classification metrics cannot capture. 

We present our experiment propagating imperfect classifications of synthetic \lcs to constraints on the dark energy equation-of-state parameter $w$ and evaluate a comprehensive set of metrics to establish how well those of classification predict those of cosmological constraining power.
This paper is organized as follows:
We review the mock \lc data set and present our adjustments made to it and the mock classification generation process in Section~\ref{sec:data}.
In Section~\ref{sec:methods}, we present the classification metrics, cosmology fitting procedures, and cosmology metrics.
We present the results of the quantitative analysis in Section~\ref{sec:results} and conclude in Section~\ref{sec:conclusion}. 
The code necessary to reproduce our results are available within the COINtoolbox\footnote{\url{https://github.com/COINtoolbox/RESSPECT_metric}} and the corresponding output data is available at Zenodo\footnote{\url{XXXX}}.

\begin{table}
    \centering
    \begin{tabular}{c|c|c|c|c|c|c}
    class   & DDF & & \% & WFD & & \%\\
    & $N$ & \% of & \salt & $N$ & \% of & \salt \Bstrut\\
    \hline
    Ia      & 8613  & 84.2      & 63.9  & 999 789   & 91.3      & 60.7 \Tstrut\\
    II      & 1028  & 10.0      & 6.3   & 72319     & 6.6       & 7.4 \\
    Iax     & 362   &  3.5      & 41.3  & 8993      & 0.8       & 14.3 \\
    Ibc     & 196   &  1.9      & 8.4   & 11603     & 1.1       & 6.7 \\
    CART    & 19    &  0.2      & 13.4  & 1136      & 0.1       & 11.9\\
    AGN     & 1     &  $<0.1$   & 0.2   & 146       & $<0.1$    & 0.1\\
    91bg    & 4     &  $<0.1$   & 0.9   & 308       & $<0.1$    & 0.8\\
    SLSN    & 4     &  $<0.1$   & 2.9   & 503       & $<0.1$    & 1.4\\
    TDE     & 1     & $<0.1$    & 1.3   & --        & --        & 0 \\
    PISN    & --    & --        & 0     & 9         & $<0.1$    & 0.8\\
    ILOT    & --    & --        & 0     & 22        & $<0.1$    & 1.3\\
    KN      & --    & --        & 0     & 1         & $<0.1$    & 0.8 \Bstrut\\
    \hline
    Total & 10 228 & 100 & & 1 094 829 & 100 & \Tstrut \Bstrut \\
    \end{tabular}
    \caption{The populations of \lcs~under each survey strategy that survive a \salt fit, as well as the survivor fraction after the \salt fit criterion.}
    \label{tab:composition}
\end{table}

\section{Data}
\label{sec:data}

We analyze cosmological parameter constraints derived from mock-classified samples of synthetic \lcs, which are described below.
Section~\ref{sec:samps} reviews the \lc data set,
Section~\ref{sec:sim} outlines how mock-classified \snia samples are created from the \lc catalog, and
Section~\ref{sec:cosmo} describes the procedures used to obtain cosmological parameter constraints.

\subsection{Light curves and distance moduli}
\label{sec:samps}

We first describe the pool of \lcs from which our cosmological samples are defined.
Section~\ref{sec:plasticc} introduces the multi-class \lcs, and Section~\ref{sec:lowz} describes the additional set of \snia \lcs included as a realistic low-redshift anchor sample.
We then present the process by which distance moduli are derived from the \lcs in Section~\ref{sec:snana}.

\subsubsection{\plasticc \lcs}
\label{sec:plasticc}

The Photometric \lsst Astronomical Time-Series Classification Challenge~\citep[\plasticc;][]{plasticc_release, Kessler19, Malz19, Hlozek21} was an open challenge that ran in 2018 and offered a cash prize to catalyze the development of \lc classifiers by the machine learning community; 
as \plasticc aimed to address a Rubin-wide need for multi-class classification, its metric was agnostic to specific science goals, leaving open the possibility of subsequent work, such as this, to explore metrics for cosmology and other use cases.
The complete unblinded \plasticc data set\footnote{\url{https://zenodo.org/record/2539456}} \citep{plasticc_data} includes simulations of three years of observations for \lsst. 

The data set was generated considering a flat dark energy dominated cosmology with dark  matter energy density $\Omega_m = 0.3$ and dark energy equation of state $w = -1$.
Fourteen galactic and extragalactic classes are represented in the training set, and 15 classes are present in the test set.  
In this work, we limit our sample to extragalactic ($z > 0$) sources in the test set, including Supernova Type Ia (SN~Ia), Supernova Iax (SN~Iax), Supernova Type Ia-91bg (SN~Ia-91bg), Core-collapse Supernova Type Ibc (SN~Ibc), Core-collapse Supernova Type II (SN~II), Super Luminous Supernova (SLSN), Tidal Disruption Event (TDE), Kilonova (KN), Active galactic nucleus (AGN), Intermediate Luminous Optical Transients (ILOT), Calcium-rich Transient (CaRT), and Pair-instability SN (PISN) models.
For more details about the \plasticc models and simulations we refer to \citet{Kessler19}.

The \plasticc simulations assumed a baseline cadence model that has two distinct observing strategies: 
Wide Fast Deep (WFD) and Deep-Drilling Fields (DDF), both covering observations in all LSST filters ($ugrizY$) and following the trigger model described in Section 6.3 of \citet{Kessler19}. 
The WFD covers 17,950 deg$^2$ every few days, producing a large set of sparsely sampled \lcs. 
The DDF covers a much smaller part of the sky and observes in at least two filters every night, yielding \lcs having higher signal-to-noise ratio (SNR) and denser time-sampling. 
In order to isolate the effect of different survey strategies on the final cosmological results, we present separate results for DDF and WFD \lcs. 

\subsubsection{Low-$z$ anchor \lc sample}
\label{sec:lowz}

We supplement all our synthetic sub-samples of \plasticc \lcs with a simulated low-redshift sample of 147 \sneia with $0.01 \leq z \leq 0.11$ as a stand-in for the common practice of supplementing photometric \sneia samples with high-fidelity, spectroscopically-confirmed \sneia. 
The simulation is generated using \snana and reproduces the \proj{Foundation} sample in \citet{jones19}.
This low-$z$ sample acts as an anchor for the Hubble diagram, thus guaranteeing numerical convergence for samples with higher contamination fractions.

\subsubsection{Distance modulus estimation}
\label{sec:snana}

We assumed the true redshifts of the host galaxies were known to avoid introducing the nonlinear bias from the \plasticc photo-$z$ model.
All \plasticc test set light curves were subjected to the SALT2 fitting and standardization procedure \citep{salt2}, and only \lcs for which this process converged were used in the subsequent analysis. 
This procedure naturally selects objects whose \lcs are similar to the SALT2 SNIa model, reducing the total number of available \lcs and raising the proportion of \snia, as detailed in Table \ref{tab:composition}. 
In effect, this not only shows that SALT2 convergence is an excellent classifier of \sneia but also that the surviving non-\snia \lcs are those with properties, e.g. absolute magnitude, most similar to \sneia.
% \revtwo{}{\footnote{}}

Subsequently, we use the \saltmu \citep{salt2mu} program within the \snana package, which uses the ``BEAMS with Bias Correction" \citep[BBC][]{Hlozek2012} method to calculate bias-corrected distance moduli of the post-SALT2 \lc sample \citep{salt2mu,KesslerScolnic2017}. 
It fits the population-level nuisance parameters $\alpha$ and $\beta$ decoupled from the cosmology fit and determines bias-correction terms by simulating large \lc samples and running them through the same analysis procedure. 
Here, we simulate the bias-correction samples using the same \snana inputs (i.e. redshift, luminosity-color, luminosity-stretch parameters) that generate the \plasticc data and the low-$z$ sample, and increase the sample size by a factor of $\sim10$. 
We then utilize a 1D bias correction method within \snana, which only determines the bias in distance modulus as a function of redshift in the context of \salt2mu (it does not calculate biases in the determination of other SALT2 parameters).

\subsection{Mock \snia classification}
\label{sec:sim}

We build the \lc samples for the cosmology calculations by considering mock classifiers of the full set of \lcs shown in Table~\ref{tab:composition}, in a procedure analogous to that of \citet{Malz19}.
In Section~\ref{sec:rates}, we describe the mock classifiers, and in Section~\ref{sec:size}, we address the choice of the size of the mock samples used in the cosmological analysis.

\subsubsection{Mock classifiers}
\label{sec:rates}

Though modern classifiers often provide classification probabilities to sophisticated \snia cosmology pipelines, the scope of our experiment only needs deterministic classifications necessary to define \lc samples.
We define three baseline mock classifiers: \per, \ran, and \fid. 
The \per classifier yields an entirely pure sample of \sneia, and the \ran classifier yields a sample with the class proportions of the underlying post-\salt \plasticc data set given in Table~\ref{tab:composition}.
The \fid classifier emulates a realistically competitive classifier, modeled after \textit{Avocado}, the winner of \plasticc \citep{Boone2019}, defined by the pseudo-confusion matrix provided in Figure 8 of \citet{Hlozek21}. 

In addition to these baseline classifiers, we construct mock classifiers with controlled levels of contamination, considering one contaminant class at a time.
To generate each \lc sample, we set a desired sample size $\mathcal{N}$ comprised of the numbers of true positives $TP$ (true \snia correctly classified as \snia) and false positives $FP$ (true non-\snia misclassified as \snia).
For a sample of $\mathcal{N} \equiv TP + FP$ \lcs classified as \snia, a fraction $c$ belong to the contaminating class, whereas the remaining $1 - c$ are true \snia. 
We consider target contamination rates of $c = 0.01, 0.02, 0.05, 0.1, 0.25$.
% \aim{Why is this equation not rendering???}
% $\left[\begin{matrix}1-c & c\\
% c & 1-c\end{matrix}\right]$
% for contamination rates $c = 0.01, 0.02, 0.05, 0.1, 0.25, 0.28$

\subsubsection{Cosmological sample size}
\label{sec:size}

Independent of the bias due to contamination, a larger cosmological sample will yield tighter constraints on the cosmological parameters.
To isolate this effect, we perform experiments with a shared sample size of $\mathcal{N} = TP + FP = 3000$ cosmological \lcs.
Though \lsst's photometric cosmology sample will be far larger, we performed a series of tests with different samples sizes which showed that keeping a sample  comparable to that of modern spectroscopic \snia cosmology analyses was enough to access the impact in cosmology we wish measure, while maintaining computational cost within feasible values for this proof of concept work. 
Nevertheless, we expect that the qualitative impact of different contaminant populations at each given contamination level will be preserved among samples of constant total size in the limit in which it is much larger than the low-$z$ anchor sample size.

As a consequence of enforcing the intrinsic balance of classes under the WFD and DDF observing strategies, some rare classes don't have enough members to draw without replacement the desired $FP = c\mathcal{N}$ non-\sneia \lcs for the contaminated samples at all target contamination rates $c$.
To preserve the realism of the test cases, we create samples only for reasonable values of $c$ given the potential pool of \lcs shown in Table~\ref{tab:composition}.
In DDF, we consider only $c$ less than or equal to the ratio of contaminant \lcs to SNIa \lcs.
Because the quality of \lcs in WFD varies so much, we perform ten trials, drawn with replacement, to establish error bars on the metrics;
we thus consider only values of $c$ less than or equal to ten times the ratio of the contaminant to SNIa in the post-\salt \plasticc sample.

\begin{table}
    \centering
    \begin{tabular}{||l||}
    \hline
    Physical Constants\\
    %\hline
    $H_0 = 70~\rm{km/s/Mpc}$\\
    $c = 3\times10^5~\rm{km/s}$\\
    \hline
    \hline
    Model Relationships\\
    %\hline
    $E(z) =\int_0^z\frac{1}{\sqrt{\Omega_m(1+z)^3 + (1-\Omega_m)(1+z)^{3(w+1)}}}$\\
    $\mu_{\rm th}(z)  =  M + 5\log_{10}\left[\frac{c}{H_0}(1+z)E(z)\right]$\\
    \hline
    \hline
    Priors\\
    %\hline
    $M \sim \mathcal{N}(0, 50)$\\
    $\Omega_m  \sim  \mathcal{N}(0.3, 0.01)$\\
    $w  \sim  \mathcal{U}(-11, 9)$\\
    \hline
    \hline
    Likelihood\\
    %\hline
   $\mu  \sim  \mathcal{N}(\mu_{\rm th}, \mu_{\rm err}^2)$\\
    \hline
    \end{tabular}
    \caption{Description of the \bayes model for cosmological parameter inference.}
    \label{tab:app}
\end{table}

\subsection{Cosmology constraints}
\label{sec:cosmo}

Using distances obtained from \code{SALT2mu}, we subject all our mock samples to a Hubble Diagram fit to obtain constraints for the dark energy equation of state parameter $w$ and the matter density parameter $\Omega_{m}$, assuming priors of $\Omega_{m}\sim\mathcal{N}(0.3, 0.01)$ and $w \sim \mathcal{U}(-11, 9)$ in all test cases. 

For comparison, we employ two approaches to constraining the cosmological parameters, the \wfit method\footnote{See SNANA manual, Section 11 at \url{https://github.com/RickKessler/SNANA/blob/master/doc/snana_manual.pdf}} implemented within \snana \citep{snana} and a simple Bayesian model for parameter inference (\bayes) which  produces full posterior estimates for $w$ and $\Omega_{m}$. 
As the tight prior on $\Omega_{m}$ dominates the joint posterior samples, we present here only the constraints on $w$. 
The \bayes model structure is given in Table~\ref{tab:app}. 

Though \bayes\footnote{\url{https://github.com/COINtoolbox/RESSPECT_metric/blob/main/utils/cosmo.stan}} does not contain many of the nuances of modern cosmology pipelines \citep[e.g.][]{Hlozek2012,KesslerScolnic2017,Hinton2020}, it is not an oversimplification given the goal of this paper.
\resspect seeks not to perform a cosmological analysis to derive physically meaningful constraints.
Rather, we aim only to quantify the effect of training set imperfection on derived cosmology results in order to identify the follow-up candidates whose inclusion in the classifier's training set will be most impactful to downstream cosmological constraints.
We thus consider a simplified cosmology pipeline and deterministic classification scenario, resulting in a conservative framework to evaluate the \revone{maximum}{potential} impact on cosmology under each case of imperfect classification. 
As our goal is to determine if \resspect needs a cosmology metric to optimally allocate spectroscopic follow-up resources for training set construction, or if classification metrics are sufficient, we require a computationally light pipeline working on incomplete data.
The framework described here is thus entirely appropriate even if it would be insufficient for a research-grade cosmological study.

\begin{figure}
    \centering
    \includegraphics[width=0.5\textwidth]{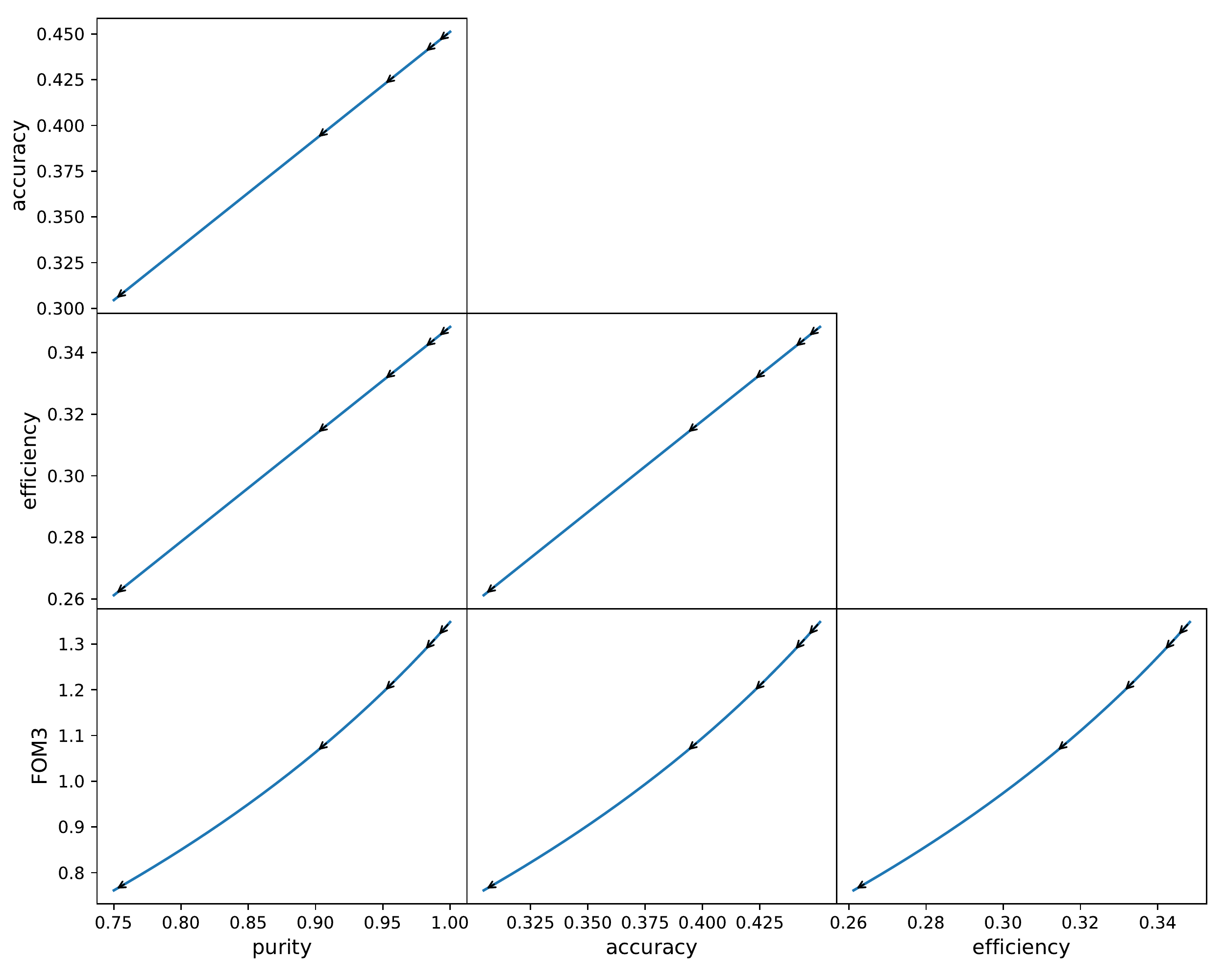}
    \caption{The traditional deterministic metrics as a function of each other for increasing values (arrows) of the contamination parameter $c=0.01, 0.02, 0.05, 0.10, 0.25$. 
    Though these metrics are functions of the same four variables (see Equations~\ref{eq:acc}, \ref{eq:pre}, \ref{eq:eff}, and \ref{eq:fom}) and should thus be expected to have consistent relationships at all values of $c$, the detailed shapes depend on the values of the true and false positive and negative rates;
    this demonstrative plot thus reflects the proportions of \sneia and non-\sneia under the DDF observing strategy from  %for the exemplary confusion matrix $\left[\begin{matrix}
%TP & FP\\
%FN & TN
%\end{matrix}\right]
%\revthree{\revtwo{}{$\propto \left[\begin{matrix}
%1-c & c\\
%c & 1-c
%\end{matrix}\right]$}}{}$ 
 Table~\ref{tab:composition} and our cosmological sample size of $\mathcal{N}=3000$ (see Section \ref{sec:size}).  
 As anticipated, these metrics are degenerate with one another and are insensitive to the contaminating class makeup, only probing the contamination rate.}
    \label{fig:det}
\end{figure}

\section{Methods}
\label{sec:methods}

We evaluate two categories of metrics: Section~\ref{sec:detmets} describes those based on the degree of non-Ia contamination within each mock cosmological sample, and Section~\ref{sec:cosmomets} describes those based on cosmological parameter constraints obtained from the same samples.

\subsection{Metrics of classification}
\label{sec:detmets}

Deterministic classifications are often summarized by a confusion matrix \citep{Hlozek21}, an array of the number of objects truly of class $i$ classified as class $j$ for all pairs $i, j = 1, \dots, M$ for $M$ classes in total.
Since the application of \snia cosmology is concerned only with the classification of \lcs as \sneia or non-\sneia, we evaluate the classification metrics in the binary case of $M=2$, i.e. \snia vs. non-\snia.
In addition to $TP$ and $FP$ defined in Section~\ref{sec:samps}, we must also define the numbers of true negatives $TN$ (true non-\snia correctly classified as non-\snia) and false negatives $FN$ (true \snia misclassified as non-\snia).
 
We evaluate the following classification metrics, initially proposed within the \snpcc \citep{Kessler10}:
\begin{itemize}
    \item The \textit{accuracy} is defined as
    \begin{equation}
    \label{eq:acc}
        \mathcal{A} = \frac{TP + TN}{\mathcal{N}},
    \end{equation}
    where a value closer to unity is more accurate.
    \item The \textit{purity} (also known as \textit{precision}) is defined as
    \begin{equation}
    \label{eq:pre}
        \mathcal{P} = \frac{TP}{TP + FP},
    \end{equation}
    where a value closer to unity is more pure.
    \item The \textit{efficiency} (also known as \textit{recall}) is defined as 
    \begin{equation}
    \label{eq:eff}
        \mathcal{R} = \frac{TP}{TP + FN},
    \end{equation}
    where a value closer to unity is more efficient.
    \item The \snpcc~ defined a \textit{Figure of Merit (FoM)},
    \begin{equation}
    \label{eq:fom}
        {\rm FoM}_{W^{\rm false}} \equiv {\rm FoM}(W^{\rm false}) = \frac{TP}{TP + FN} \times \frac{TP}{TP + W^{\rm false} \times FP},
    \end{equation}
    where the factor $W^{\rm false}$ penalizes false positives. For $W^{\rm false} = 1$, $FoM_{1} = \mathcal{R} \times \mathcal{P}$.
    We use $FoM_{3}$ in this paper to match the \snpcc value of $W^{\rm false} = 3$.
\end{itemize}

Figure~\ref{fig:det} shows the aforementioned metrics as a function of contamination parameter $c$, showing that they are wholly degenerate with one another and insensitive to the contaminant types.
As a consequence, we only need to evaluate one classification metric and choose $FoM_{3}$, noting that in our experimental design $TN = FN = 0$.

\subsection{Metrics of cosmology constraints}
\label{sec:cosmomets}

We explore metrics of derived cosmological constraints between our synthetic \snia samples, rather than relative to an absolute true cosmology.
By doing this, we account for the fact that the purity of the \snia sample is not the only factor influencing the quality of the cosmology results;
the quality of the light curves themselves, for example, and the analysis methodology chosen both impact the accuracy and precision of derived constraints.
This paper aims to isolate such effects from that of systematic deviations from a perfect classification.

We compare cosmology metrics that can be divided into three broad categories:
a Fisher Matrix based on redshifts and distance moduli under the \lcdm cosmological model,
a Gaussian approximation to the inferred $w$,
and measures of the inferred posterior probability distribution of $w$, each relative to that of the \per sample Section \ref{sec:rates}).

\begin{itemize}
\item The \textit{Fisher Matrix (FM)} from \lc fits: 
    Frequently used to guide survey design decisions, the Fisher Matrix uses redshifts and estimated errors on distance moduli under a Gaussian likelihood centered on a given mean model \citep{Albrecht06}, probing only the expected uncertainties in inferred parameters. 
    We calculate the FM at the expected mean of $w=-1$ and $\Omega_{m} = 0.3$ given a flat universe and report the fractional difference $\Delta FM$ on the inverse of its diagonal component $\sigma^{2}_{w}$ between a given \lc sample's estimate and that of the \per\ sample.

\item The summary statistics of estimated cosmological parameters:
     \wfit assumes a Gaussian likelihood centered on the $\Lambda$CDM model and produces an estimated \textit{mean} $\hat{\mu}_{\wfit}$ and \textit{standard deviation} $\hat{\sigma}_{\wfit}$ whereas \bayes, on the other hand, yields posterior samples of $w$, which define a univariate probability density function (PDF). 
     For the sake of comparison, we fit a normal distribution to the \bayes posterior samples of $w$ to obtain $\hat{\mu}_{\bayes}$ and $\hat{\sigma}_{\bayes}$ and observe their relative response under different contamination levels and contaminant classes.

\item Metrics of cosmology posterior PDFs:
    The posterior samples of $w$ from the \bayes fit define a PDF, which we flexibly fit and evaluate on a fine grid using kernel density estimation (KDE), i.e.~eliminating the Gaussian assumption of the aforementioned cosmology metrics.
    We then perform a quantitative comparison of the KDEs $\hat{p}_{mock}(w)$ for each synthetic \lc sample by comparing them to that of the \per sample $\hat{p}_{0}(w)$ by evaluating two metrics:
\begin{itemize}
\item The \textit{Kullback-Leibler Divergence (KLD)},
\begin{equation}
    \label{eqn:kld}
    KLD = -\int \hat{p}_{0}(w) \ln\left[ \frac{\hat{p}_{mock}(w)}{\hat{p}_{0}(w)} \right] dw,
\end{equation}
is an information theoretic measure of the loss of information due to using an approximation $\hat{p}_{mock}(w)$ rather than the true distribution $\hat{p}_{0}(w)$;
the KLD has been used before in extragalactic astrophysics \citep{Malz18,2020ApJ...890...74K}.

\item The \textit{Earth-Mover's Distance (EMD)}
\begin{equation}
\label{eqn:emd}
    EMD = \int_{-\infty}^{\infty} \left| \int_{-\infty}^{w} \hat{p}_{0}(w') dw' - \int_{-\infty}^{w} \hat{p}_{mock}(w') dw' \right| dw, 
\end{equation}
also known as the first order \textit{Wasserstein metric}, can be intuitively understood as the integrated discrepancy between a pair of PDFs, defined in terms of their cumulative distribution functions (CDFs);
the EMD has been used before in cosmology \citep[e.g.][]{Moews21}.
\end{itemize}
For both the KLD and EMD, lower values indicate a closer correspondence between distributions.
\end{itemize}
%--------------------------------------------------------------------

\begin{figure}
    \centering
    \includegraphics[scale=0.46]{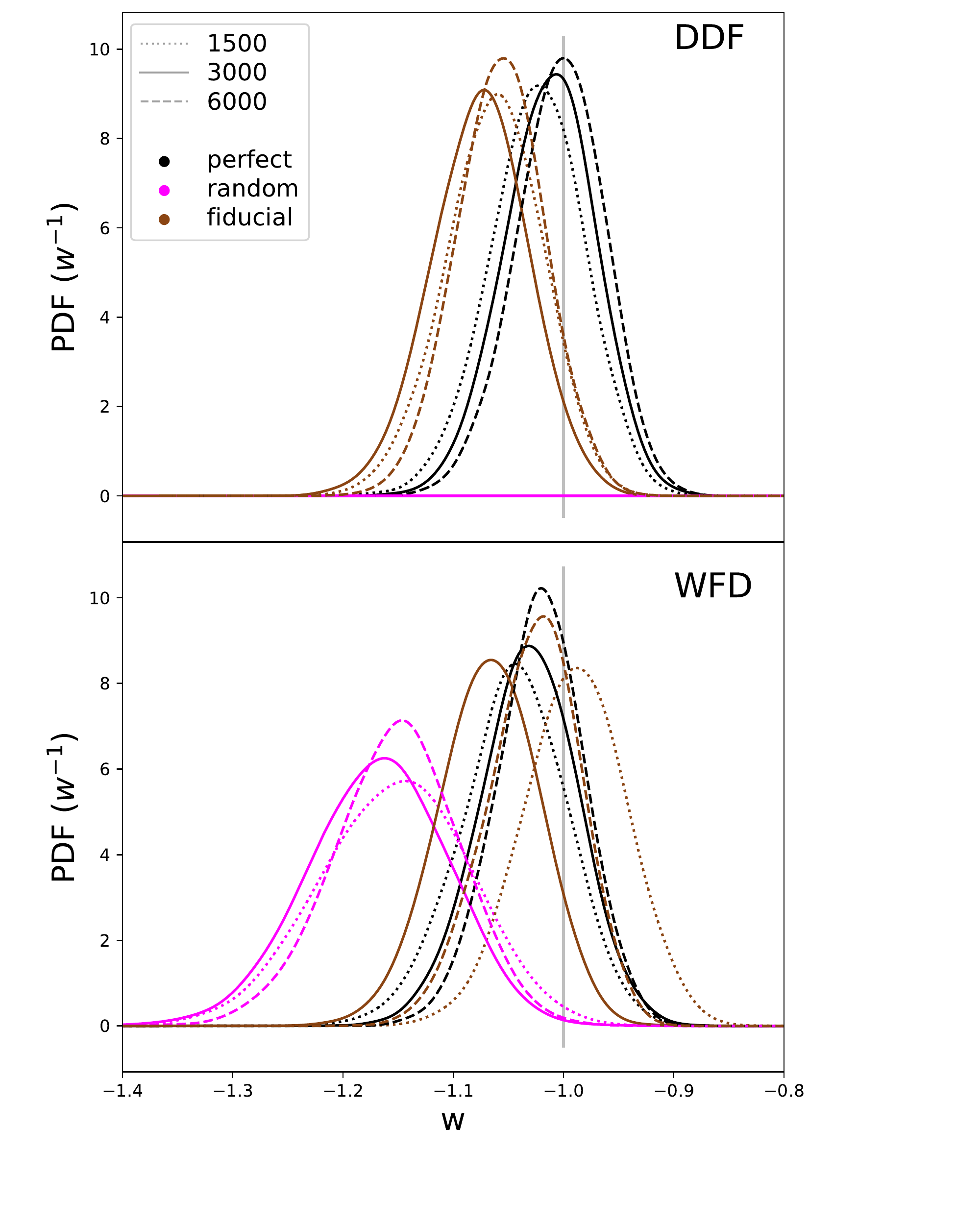}
    \caption{
    \bayes posterior PDFs of $w$ for different cosmological sample sizes (line styles) from the DDF and WFD observing strategies (panels) under the \per~(black), \ran~(magenta), and \fid~(brown) mock classification schemes; 
    note that the random classifier's constraints in the DDF underestimate $w$ by so much that they cannot be shown on these axes.
    As the constraints are not very sensitive to the sample size, we can safely use a cosmological sample of 3000 \lcs in our tests. 
    }
    \label{fig:contour}
\end{figure}

\begin{figure*}
    \centering
    \includegraphics[width=\textwidth]{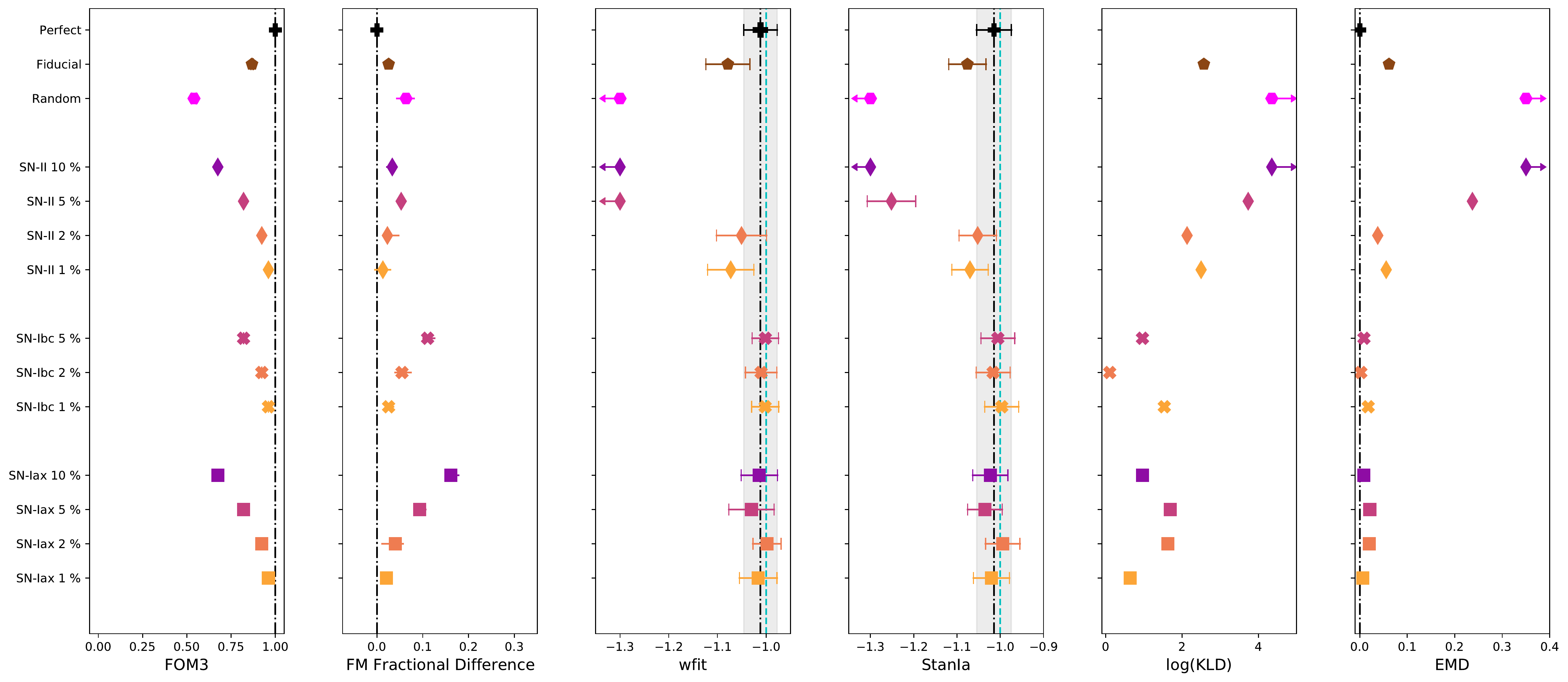}
    \caption{Metrics summary for the DDF, with metric values on the x-axes and \lc samples on the y-axis, grouped by contaminant (shape and light purple backgrounds) and ranked by contamination fraction (color) aside from the \per, \fid, and \ran~\lc samples defined in Section~\ref{sec:sim}. 
    Reference values (vertical lines; dotted black for the pure sample with 1-$\sigma$ error regions in gray and dashed cyan for the truth) are provided where appropriate. 
    The constraints on $w$ (central two panels) include both the mean $\hat{w}$ and standard deviation $\sigma^{2}_{w}$.
    }
    \label{fig:wdiff_ddf}
\end{figure*}

\begin{figure*}
    \centering
    \includegraphics[width=\textwidth]{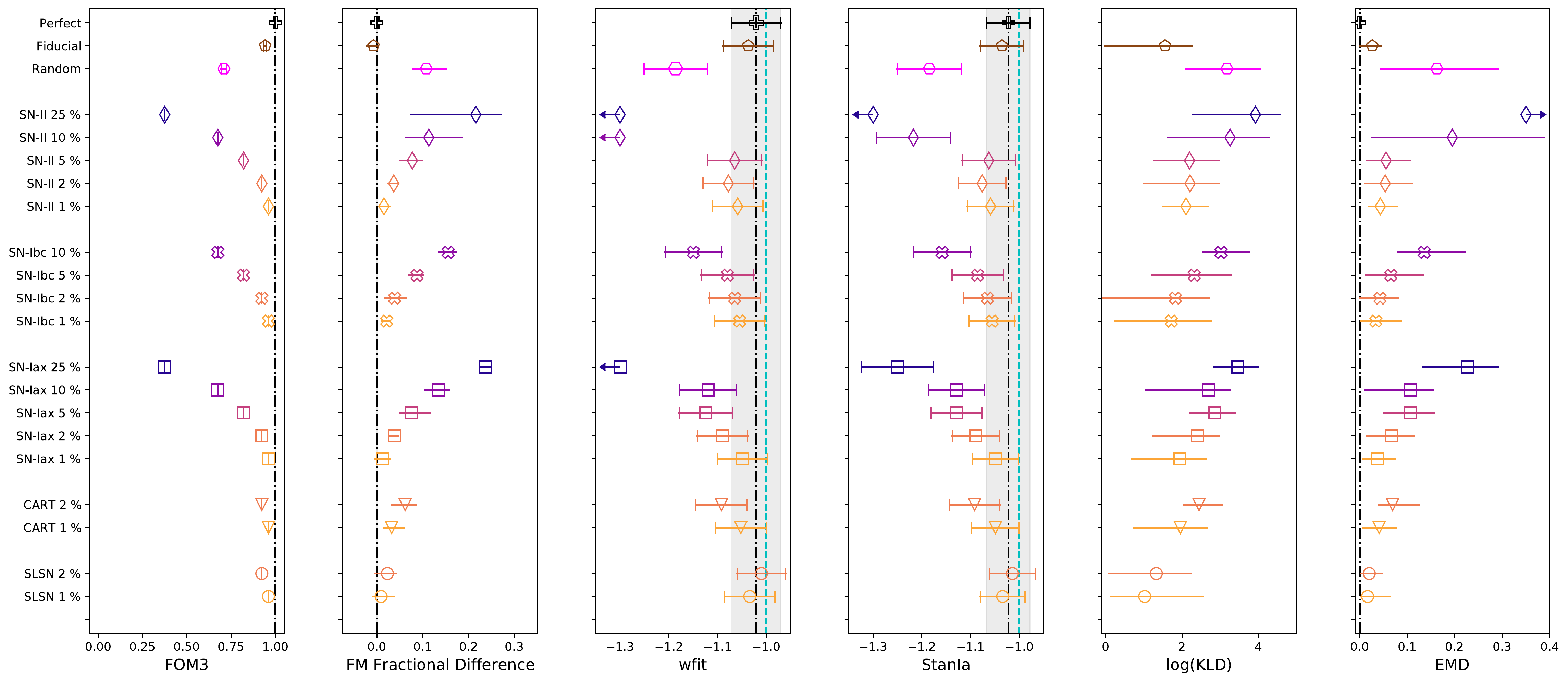}
    \caption{Equivalent of Figure \ref{fig:wdiff_ddf} for the WFD based on ten realizations of each sample.
    The plotted uncertainties in the constraints on $w$ (central two panels) correspond to the largest $\sigma^{2}_{w}$ out of the ten trials.
    Similarly, the plotted uncertainties on the posterior PDFs on $w$ (rightmost two panels) and the Fisher Matrix fractional difference (second panel from left) indicate the maximum and minimum metric values out of the ten trials.
    }
    \label{fig:wdiff_wfd}
\end{figure*}

\begin{figure*}
    \centering
    \noindent\includegraphics[width=0.9\textwidth]{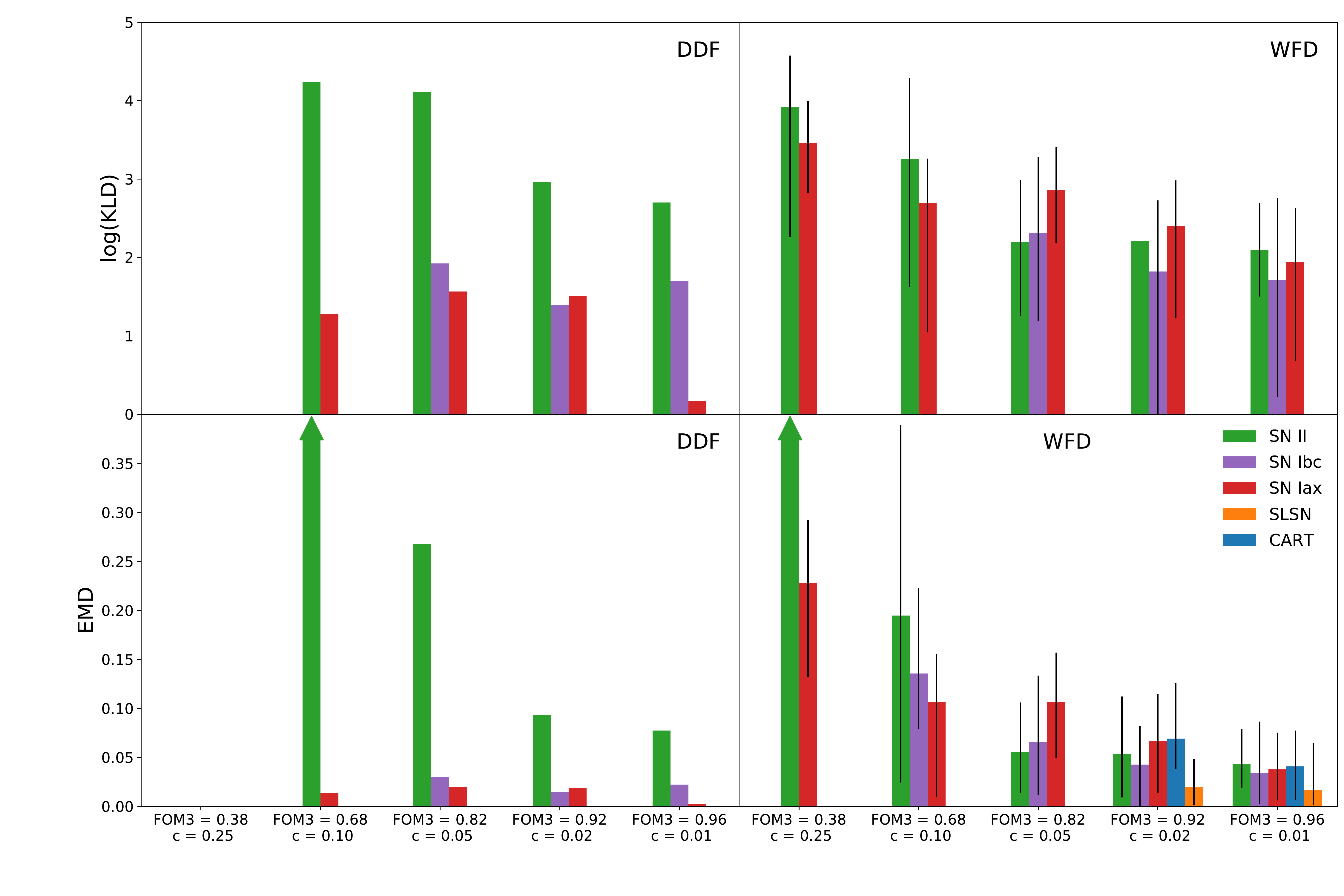}
    \caption{The log-KLD and EMD of posterior samples of $w$ at different values of the \fomthree classification metric for each contaminating class (color) and observing strategy (panel). 
    The vertical lines within the WFD panels indicate the minimum and maximum values among ten different realizations.
    Despite the considerable variability between realizations, the cosmology metrics do exhibit consistent sensitivity to the impact of different classes of contaminant at a given contamination level.
    Examples of this effect manifest as stratification among the classes at a given \fomthree value.
    }
    \label{fig:kld}
\end{figure*}

\section{Analysis \& Results}
\label{sec:results}

Recalling that the goal of this investigation is to assess the degree to which classification metrics are consistent with metrics of cosmological constraints in the context of \resspect's need for an internal metric to optimize classifications ultimately destined for \snia cosmology, we now present a comprehensive comparison of various metrics evaluated on incrementally contaminated samples.

Our first goal is to quantify the effect on parameter inference due to sample size. 
Figure~\ref{fig:contour} shows posterior samples of $w$ for the \per, \ran, and \fid cases on mock cosmological \lc samples for different sample sizes.
As the observed sensitivity of the posterior PDFs on $w$ to sample size matches intuitive predictions (i.e.~narrower for larger sample size), it is thus safe to use $TP + FP = 3000$ ``post-classification" \lcs in our cosmological samples. 
The relatively small difference between the posterior widths for the DDF and WFD \lcs could be considered a natural consequence of the fact that the samples include only \lcs that survived a \salt fit and thus have comparably-sized error bars on the distance moduli that enter the cosmology fits.

The tremendous gap between \ran and the other two samples in the DDF is a direct consequence of the intrinsically higher signal-to-noise ratio and sampling rate defining the DDF observing strategy. 
We also see that under DDF conditions, the constraints of the realistic \fid sample are close to the results for the \per sample. 
For the WFD, however, the distinction between the three cases is less pronounced, although the \ran sample still outputs the largest biases independent of the sample size. 

Figures~\ref{fig:wdiff_ddf} and \ref{fig:wdiff_wfd} depict the behavior of our metrics as a function of the contaminant class and contamination level for the DDF and WFD observing strategies, respectively.
Under both observing strategies, we note that the metrics of derived cosmological constraints are sensitive to both the contaminant class and the contamination rate, whereas the classification metric probes only the rate.

Under the DDF, we observe that $\Delta FM$ shows a comparable impact for 2\% SN II and 1\% SN Ibc, and, separately for 5\% SN II and 2\% SN Ibc contamination, which are themselves on par with that of the random classification scheme, indicating that SN Ibc contaminants are effectively twice as damaging as SN II contaminants, and that random contamination isn't much worse than that.
However, the metrics derived from a full cosmological analysis tell a different story;
the constraints from \wfit and \bayes agree that even 1\% contamination with SN II skews the mean $\hat{w}$ beyond the 1-$\sigma$ error bars of the pure sample, whereas even 5\% SN Ibc or SN Iax do not.
Critically, the bias due to even a low contamination rate by SN II is on par with what would result from the realistic \fid classifier, a concern mirrored in the response of the metrics of the posterior PDFs of $w$ from \bayes.

\begin{figure*}
    \centering
    \includegraphics[width=\textwidth]{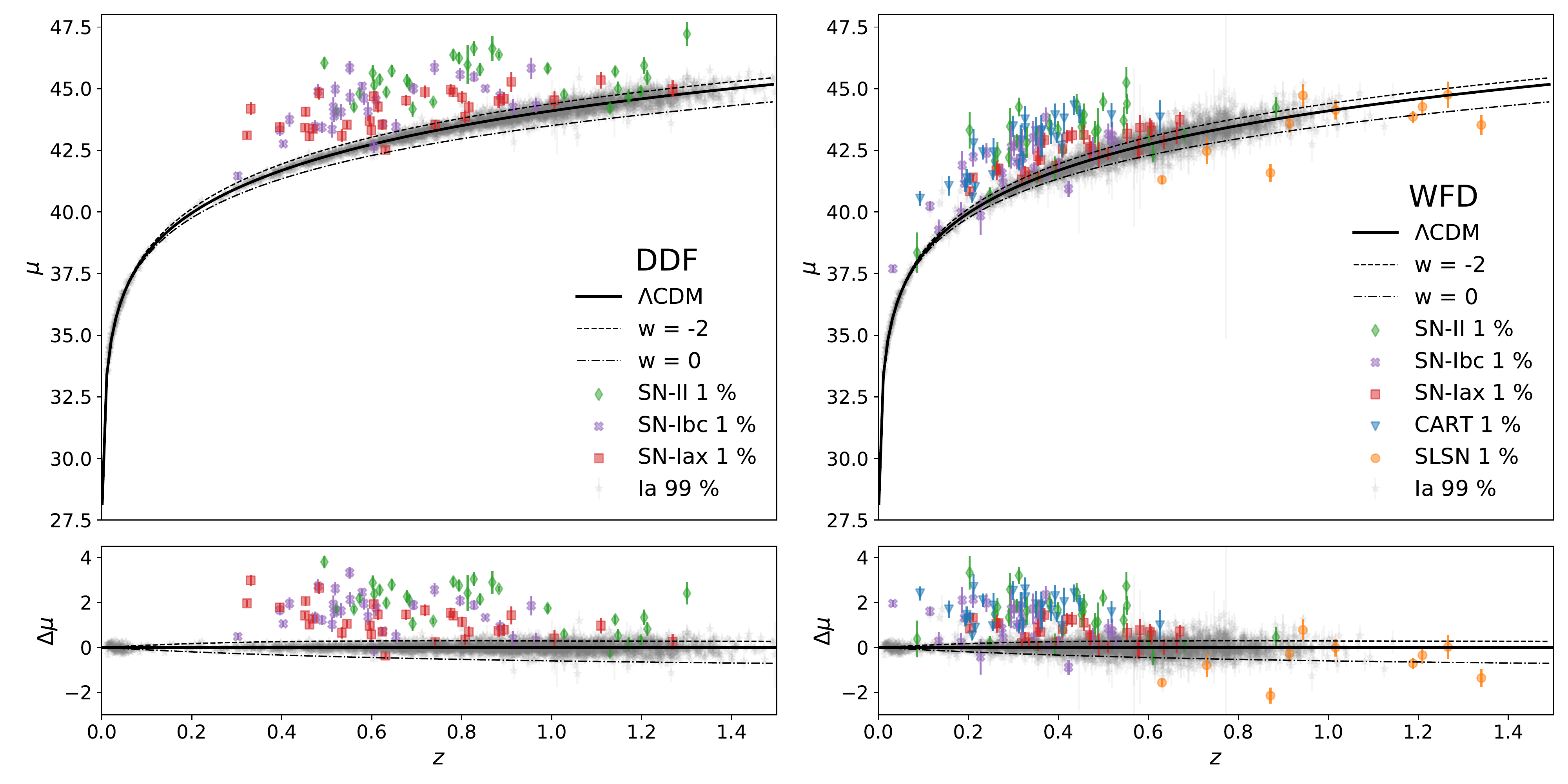}
    \caption{Hubble diagram (upper) and residuals (lower) showing the \lcdm model (solid black line) and two others depicting extreme dark energy behaviors ($w = -2$ - dashed, and $w = 0$ -- dot-dashed). 
    The gray points correspond to 2970 randomly selected \sneia from the available sample plus the low-$z$ anchor \sneia, and the contaminants are shown at 1\% contamination (class-specific shapes and colors) in DDF (left) and WFD (right). 
    While some individual contaminant \lcs have \saltmu fit parameters far from those of the true \sneia, there is nontrivial overlap that would preclude simply classifying by eye to remove them from a sample entirely as expected given the selection criterion of a convergent \saltmu fit);
    in the DDF, this effect is noticeable among the SN-Iax contaminants as well as SN-Ibc and SN-II at $z > 0.8$,
    whereas in the WFD, the problem is more severe, affecting all contaminants except CART and at redshifts as low as $z > 0.2$.
    }
    \label{fig:hubble}
\end{figure*}

For visualization purposes, Figure~\ref{fig:wdiff_wfd} displays error bars corresponding to the greatest deviation from the mean across the ten trials rather than the standard deviation, except for the \fomthree metric, which lacks error bars because it is the same across all trials by construction.
The most striking effect in Figure~\ref{fig:wdiff_wfd} is that the variation in metric values due to the random sample of included \lcs dominates over the impact of the different contaminant identities;
there is a large range of \lc quality under the WFD observing strategy, and our relaxed sample selection criteria permit what amounts to only a few \lcs to sway the cosmological constraints.
Beyond that, we observe that 1\% and 2\% contamination by all classes are indistinguishable by all cosmology metrics and do not induce a bias inconsistent with a pure sample nor the \fid mock classifier, a reassuring discovery.
Though there is a weakly class-dependent effect at higher contamination rates according to the estimated mean and standard deviation on $w$ by both fitting methods, which shows that 5\% contamination with SN Iax is worse than 5\% contamination by SN Ibc or SN II, the effect only persists at 10\% contamination for \wfit and at 25\% for \bayes, suggesting a need for more trials.

Figure~\ref{fig:kld} directly compares the relative response of the \fomthree classification metric and the KLD and EMD of posterior samples of the cosmological parameters for subsamples of varying contamination rate and contaminant within the DDF and WFD observing strategies.
The clustering of points at discrete values of \fomthree are a result of its insensitivity to contaminant class, and the differentiation within each group demonstrates the sensitivity of the resulting cosmological parameter constraints to the type of contaminant at the same contamination rate.
As is observed in Figures~\ref{fig:wdiff_ddf} and \ref{fig:wdiff_wfd}, we see stratification of cosmology metric values by contaminant class, somewhat suppressed in the WFD.
This visualization more directly highlights the conclusions from Figures~\ref{fig:wdiff_ddf} and \ref{fig:wdiff_wfd}, that at constant contamination rate, there are systematic, quantifiable differences in the derived cosmology depending on the contaminant class; 
that the effect establishes that contamination by SN II more strongly impacts the derived cosmology; 
that the variation between contaminant class is subdominant to the quality of the \lcs under each observing strategy;
and that both metrics of posterior samples of $w$ qualitatively agree.

Figure \ref{fig:hubble} shows that the severity of bias in the estimated cosmological parameters as a function of contaminant class is also related to how far off the estimated distance moduli are from the truth when fitting non-SN Ia with the SN Ia standardization model, as expected.
More importantly, it shows that individual contaminating \lcs cannot, in general, be isolated from the \snia sample based on their fitted absolute magnitude, particularly at higher redshifts and under the WFD observing strategy.
In effect, our mock sample generation procedure probes the most extreme bias that could be caused by each contaminant class. 
This test effectively includes redshift-dependent misclassification, which would lead to more of the brightest contaminants at higher redshift and those most similar to \snia in lower redshifts, thus imposing a more subtle bias in the cosmological parameter constraints that would nonetheless not be reflected in the classification metrics alone.

\section{Conclusions}
\label{sec:conclusion}

Metrics of \sneia classification often serve as proxies for metrics of cosmological constraints derived from samples of \lcs classified as \sneia, particularly in applications assessing the performance of \lc classifiers intended for cosmological analyses.
In this work, we test the strength of the assumption underlying this usage and find that classification metrics are not always an appropriate substitute for metrics of cosmological parameters;
the metrics of cosmological constraining power are sensitive to the composition of the contaminating populations as well as the contamination rate, but only the latter is probed by classification metrics.
We thus recommend the use of cosmology-based metrics in place of classification metrics when optimizing analysis pipeline designs, despite their associated computational expense, except when the \lcs are noise-dominated. 

In the context of \resspect, the above results confirm that relevant information is encapsulated in a metric of impact on cosmological constraints and should thus be considered as a factor in selecting spectroscopic follow-up candidates for inclusion in the training set within the active learning pipeline.
More generically, as astronomical classifications are of course used for many other population-level studies including and beyond transients, we encourage a healthy skepticism to those aiming to use such classifications in further scientific analyses;
it would be prudent to confirm any correspondence between classification performance and metrics tailored to a specific science case prior to any decision-making on analysis approaches.

{\small
\subsubsection*{Author Contributions}
% %Conceptualization
% %Data curation
% %Formal analysis
% %Funding acquisition
% %Investigation
% %Methodology
% %Project administration
% %Resources
% %Software
% %Supervision
% %Validation
% %Visualization
% %Writing - original draft.
% %Writing - review & editing
A.I.~Malz: conceptualization, formal analysis, investigation, methodology, software, validation, visualization, writing - original draft, writing - review \& editing\\
M.~Dai: data curation, formal analysis, investigation, methodology, software, validation, writing - review \& editing\\%Conceptualization, Formal analysis, investigation, methodology, software, writing - review and editing% Conceptualization, Data curation, Software, Validation, Writing - review & editing
K.A.~Ponder: conceptualization, formal analysis, investigation, methodology, software, validation, visualization, writing - original draft\\%Conceptualization, formal analysis, investigation, methodology, visualization, writing - original draft
E.E.O.~Ishida: conceptualization, data curation, formal analysis, funding acquisition, project administration, resources, software, supervision, validation, visualization, writing - original draft, writing - review \& editing\\%conceptualization, formal analysis, funding acquisition, project administration, software, writing - original draft
S.~Gonzalez-Gaitain: conceptualization, methodology, software, writing - review \& editing\\%conceptualization, methodology, software%Writing - review & editing
R.~Durgesh: software\\%software%Software
A.~Krone-Martins: funding acquisition, project administration, resources, software, supervision\\%conceptualization, funding acquisition, resources
R.S.~de Souza: funding acquisition, project administration, resources, supervision\\%conceptualization, funding acquisition, resources
N.~Kennamer: software, methodology\\%software, methodology
S.~Sreejith: software\\%software - writing - review and editing
L.~Galbany: conceptualization\\%conceptualization%Conceptualization, Methodology, review & editing
}

\begin{acknowledgements}
      This paper has undergone internal review in the LSST Dark Energy Science Collaboration. 
      The authors would like to thank Ren\'ee Hlozek, Alex Kim, and Maria Vincenzi for serving as the LSST-DESC publication review committee, as wel as David O. Jones, for their comments and suggestions that improved the quality of this manuscript. \\
    
      AIM acknowledges support during this work from the Max Planck Society and the Alexander von Humboldt Foundation in the framework of the Max Planck-Humboldt Research Award endowed by the Federal Ministry of Education and Research.
      AIM is a member of the LSST Interdisciplinary Network for Collaboration and Computing (LINCC) Frameworks team;
      LINCC Frameworks is supported by Schmidt Futures, a philanthropic initiative founded by Eric and Wendy Schmidt, as part of the Virtual Institute of Astrophysics (VIA).
      M.D. is supported by the Horizon Fellowship at the Johns Hopkins University.  
      S.G.G. acknowledges support by FCT under Project CRISP PTDC/FIS-AST-31546/2017 and UIDB/00099/2020.
      L.G. acknowledges financial support from the Spanish Ministry of Science and Innovation (MCIN) under the 2019 Ramon y Cajal program RYC2019-027683 and from the Spanish MCIN project HOSTFLOWS PID2020-115253GA-I00. \\
      
      This work is financially supported by CNRS as part of its MOMENTUM programme under the project \textit{Adaptive Learning for Large Scale Sky Surveys}. 
      The Cosmostatistics Initiative (COIN, \url{https://cosmostatistics-initiative.org/}) is an international network of researchers whose goal is to foster interdisciplinarity inspired by Astronomy.\\

      The DESC acknowledges ongoing support from the Institut National de Physique Nucl\'eaire et de Physique des Particules in France; 
      the Science \& Technology Facilities Council in the United Kingdom; 
      and the Department of Energy, the National Science Foundation, and the LSST Corporation in the United States.  
      DESC uses resources of the IN2P3 Computing Center (CC-IN2P3--Lyon/Villeurbanne - France) funded by the Centre National de la Recherche Scientifique; 
      the National Energy Research Scientific Computing Center, a DOE Office of Science User Facility supported by the Office of Science of the U.S.\ Department of Energy under Contract No.\ DE-AC02-05CH11231; 
      STFC DiRAC HPC Facilities, funded by UK BIS National E-infrastructure capital grants; and the UK particle physics grid, supported by the GridPP Collaboration.  
      This work was performed in part under DOE Contract DE-AC02-76SF00515.

\end{acknowledgements}

\bibliographystyle{aa} % style aa.bst
\bibliography{myrefs} % your references Yourfile.bib

\end{document}